\documentclass[11pt,twoside
]{article}

\usepackage{jae2016} 
\usepackage{graphicx}
\usepackage{subfigure}
\usepackage{psfrag}
\usepackage{amssymb}
\usepackage[spanish,activeacute,english]{babel}
\usepackage[latin1]{inputenc}
\usepackage[T1]{fontenc} 
\usepackage{ae,aecompl} 
\usepackage{latexsym}
\usepackage{verbatim}
\usepackage{amsmath}
\usepackage{stmaryrd}
\usepackage{amsfonts}
\usepackage{amssymb}
\usepackage{wasysym}

\begin{document}
\myselectenglish
\vskip 1.0cm
\markboth{A. H. C\'orsico}%
{Pulsating white dwarfs and asteroseismology}

\pagestyle{myheadings}

\title{Pulsating white dwarf stars and asteroseismology}

\author{A. H. C\'orsico$^{1,2}$}

\affil{%
  (1) IALP (CCT La Plata, CONICET-UNLP)\\
  (2) Facultad de Ciencias Astron\'omicas y Geof\'isicas-UNLP\\
}

\begin{resumen}
Actualmente,  un gran n\'umero de estrellas enanas blancas  variables
pulsantes est\'a siendo decubierto   ya sea a partir de  relevamientos
basados en observaciones desde  Tierra, tales como el Sloan Digital
Sky Survey (SDSS), u observaciones  desde el espacio (por ejemplo, la
Misi\'on Kepler). Las t\'ecnicas  astrosism\'ologicas
permiten inferir detalles de la  estratificaci\'on
qu\'imica interna, la masa total, e  incluso el perfil de rotaci\'on
estelar. En este trabajo, describimos  en primer lugar las propiedades
b\'asicas de las estrellas enanas blancas y  sus pulsaciones, as\'i
como tambi\'en los diferentes sub-tipos de estas  variables conocidos
hasta el momento. Posteriormente, describimos algunos hallazgos
recientes en relaci\'on a enanas blancas pulsantes de baja masa.
\end{resumen}

\begin{abstract} 
At present, a large number of pulsating white dwarf (WD) stars is being
discovered either from Earth-based surveys  such as the Sloan Digital
Sky Survey, or through observations from space (e.g., the Kepler
mission). The asteroseismological techniques allow us to infer details of
internal chemical stratification, the total mass, and even the stellar
rotation profile. In this paper, we first describe the basic
properties of WD stars and their pulsations, as well as the
different sub-types of these variables known so far. Subsequently,  we
describe some recent findings about pulsating low-mass WDs.
\end{abstract}

\section{WD stars in a nutshell}

WD stars constitute the ultimate fate 
of most the stars that populate the Universe, including 
our Sun. Indeed, single stars that born with masses below
$\sim 10.6 M_{\odot}$ (Woosley \& Heger 2015) will end their lives 
as WDs. Typical ages of WDs are in the range $1-10$ Gyr
(1 Gyr $\equiv 10^{9}$ yr).
Details about the formation and evolution of WDs can be found 
in the review papers of Winget \& Kepler (2008), 
Fontaine \& Brassard (2008), and Althaus et al. (2010). 
Here, we only give a brief summary of the main characteristics 
of these stars. WDs are characterized by stellar masses comparable to 
the mass of the Sun ($M_{\star} \sim 0.15-1.2 M_{\odot}$) but with 
sizes of the order of the size of planets ($R_{\star} \sim 0.01 R_{\odot}$),
which implies very high mean densities ($\overline{\rho} 
\sim 10^6$ gr/cm$^3$) and strong degeneration of matter.
In particular, the equation of state describing the
mechanical properties of WDs is that of a Fermi gas of degenerate electrons
(Chandrasekhar 1939),
which provides most of the pressure that counteracts the force of
gravity, thus preventing the collapse. Immersed in this sea of electrons ---and
decoupled thereof--- coexists a non-degenerate gas of ions that
provide the heat reservoir of the star, resulting from the previous
evolutionary history.

WDs cover a wide range of effective temperatures 
($4000 \lesssim T_{\rm eff} \lesssim 200\,000$ K) and 
hence a large interval of luminosities 
($0.0001 \lesssim L_{\star}/L_{\odot} \lesssim 1000$).
Average mass WDs ($M_{\star} \sim 0.6 M_{\odot}$) harbor
cores likely made of $^{12}$C and $^{16}$O, although  massive WDs
could have cores made of $^{16}$O, $^{20}$Ne y $^{24}$Mg,
and the lowest-mass WDs (extremely low mass WDs, abbreviated as ELM WDs)
could contain cores of $^{4}$He. Roughly speaking, WD evolution
is nothing but a slow cooling (Mestel 1952), in which the star
gets rid of its thermal
energy content into space, being the nuclear energy sources (nuclear burning)
almost extinct. The rate at which this heat is removed
(i.e., the cooling rate of the WD) is controlled by the outer
layers of the star. According to the chemical composition of the outer
layers, WDs come in two main flavors: DAs ($80 \%$, almost pure-H atmospheres)
and DBs ($15 \%$, almost pure-He atmospheres), although there are also
some WDs which show atmospheres of He, C, and O (PG1159 stars) or
C and He (DQ WDs). We have to add to this list the DZ WDs,
which show only metal lines, without H or He present. Finally,
there is the exotic object called SDSSJ1240+6710, that has
an almost pure O atmosphere, diluted only by traces of Ne,
Mg, and Si (Kepler et al. 2016a). In the case of DA
and DB WDs, the purity
in the surface chemical composition is the result of the
extremely high surface gravity ($\log g \sim 6-9$ [cm/s$^2$]),
which results in the settling of the heavier elements, leaving the
lightest nuclear species to float at the stellar surface. 

\begin{figure}[!ht]
  \centering
  \includegraphics[width=0.80\textwidth]{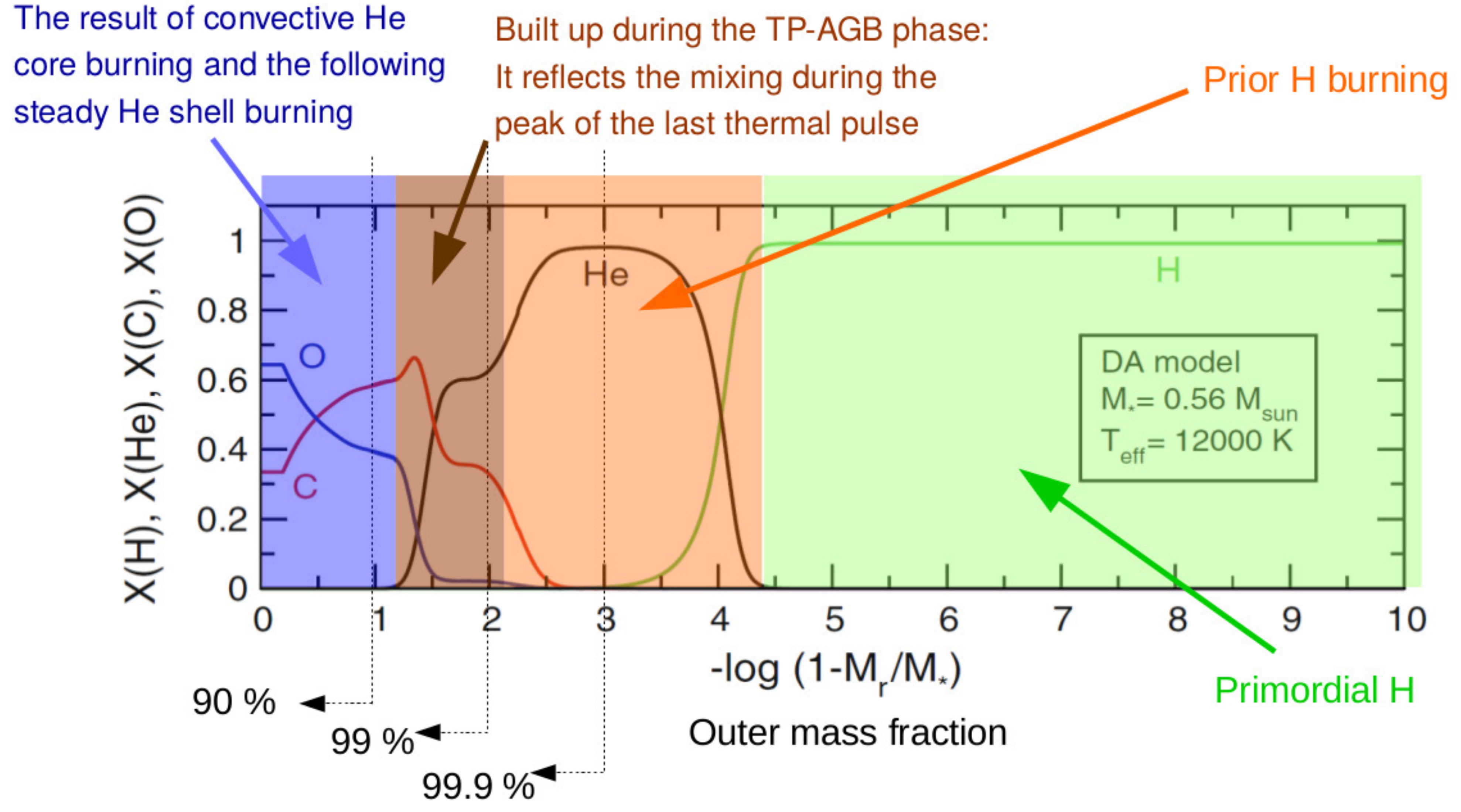}
  \caption{The internal chemical structure of a DA WD model with 
  $M_{\star}= 0.56 M_{\odot}$ and  $T_{\rm eff}\sim 12\,000$ K. Plotted
    is the mass fraction ($X$) of $^{16}$O, $^{12}$C, $^4$He, and $^1$H
    in terms of the outer mass coordinate. We include in the plot
    a brief explanation about the origin of each part of the
    internal chemical structure. Also, we indicate a few values of
    the percentage of the total mass of the star associated to
    the outer mass coordinate.} 
  \label{fig:1}
\end{figure}

In Fig. \ref{fig:1} we plot the internal chemical structure of a
typical DA WD model with $M_{\star}= 0.56 M_{\odot}$ and
$T_{\rm eff}\sim 12\,000$ K. For instructional purposes,
we have included in the figure a brief
explanation indicating the origin of each feature in the
chemical structure of the star.  Note that the adopted $x$-coordinate,
$-\log(1-M_r/M_{\star})$, strongly amplifies the outer part of the
star. As it can be seen in the figure, in a typical WD star
the $\sim 99 \%$ of the mass is
made of a mixture of $^{12}$C and $^{16}$O in uncertain proportions,
being the $^4$He content of $M_{\rm He} \sim 0.01 M_{\star}$ at most,
and the $^1$H content at the envelope of
$M_{\rm H} \lesssim 0.0001 M_{\star}$.

The number of known WD stars is increasing fast thanks to the
Sloan Digital Sky Survey (SDSS;  York  et  al.
2000). In fact, SDSS  increased
the  number  of spectroscopically-confirmed   WD  stars  more
than an order of magnitude prior to the SDSS (see Kepler et al. 2016b).
At present, there is a total of around $37\,000$ WDs identified,
$\sim 30\,000$ of which are from DR7 (Kleinman et al. 2013),
DR10 (Kepler et al. 2015) and DR12 (Kepler et al 2016b) of
SDSS, and $\sim 5000$ correspond to the
McCook \& Sion (1999) catalog
(S. O. Kepler, private communication).

\section{Why do we care about WDs?}

WDs are very old objects, so they harbor valuable information about
the history of our Galaxy. As such, these exotic objects have
application to various fields of modern astrophysics:

\begin{itemize}

\item[-] Cosmochronology: the derivation of the ages of stellar
populations using the WD luminosity function. WDs provide independent
chronometers for the age and star formation history of the Galactic
disk (e.g., Harris et al. 2006), and halo (Isern et al. 1998) through
cosmochronology.  Furthermore, WD cosmochronology can be applied to
Galactic globular clusters (Hansen et al. 2013) and open clusters
(e.g., Garc\'ia-Berro et al. 2010) as well.

\item[-] Initial to Final Mass Relation (IFMR; Catal\'an et al. 2008): more than 95 \%
of stars will end their lives as WDs, so that they can be considered
as a boundary condition for stellar evolution. The difference between
the initial Main Sequence masses and the final WD masses highlights
the role of mass-loss during the AGB phase.

\item[-] Progenitors of SNIa, cataclysmic variables (novae): energetic events
($10^{44}-10^{51}$ erg) of mass transfer on the WD by its companion.
  
\item[-] Cosmic laboratories: the interiors of WDs are composed of
matter under extreme conditions (that is, extremely high density and
pressure). This allows the study of equation of state (EoS), crystallization,
particle physics, fundamental constant variations.

\end{itemize}

Until recent years, the only available tools for studying WDs were
spectroscopy, that provides $T_{\rm eff}$, $\log g$, the surface chemical
composition, the magnitude of magnetic fields, and  photometry, that
allows to infer the stellar mass. These techniques
provide global information of the WDs. But, in what way
could we dig beneath the surface layers of these stars?  WD
asteroseismology, that is, the comparison of the observed pulsational
spectrum of variable WDs with that emerging from theoretical models
of evolution and pulsations, comes to help (Winget \& Kepler 2008;
Fontaine \& Brassard 2008; Althaus et al. 2010).
Asteroseismology allows to ``see'' inside these stars,
otherwise inaccessible by other means
(Catelan \& Smith 2015).  WD
asteroseismology allows to infer the chemical stratification, the
core chemical composition, and the stellar masses, to measure rotation and
magnetic fields, to detect planets around WDs, and to probe for
exotic particles that are strong candidates for dark matter, like
axions. 

\section{A little bit about non-radial WD pulsations}

\begin{figure}[!ht]
  \centering
  \includegraphics[width=0.8\textwidth]{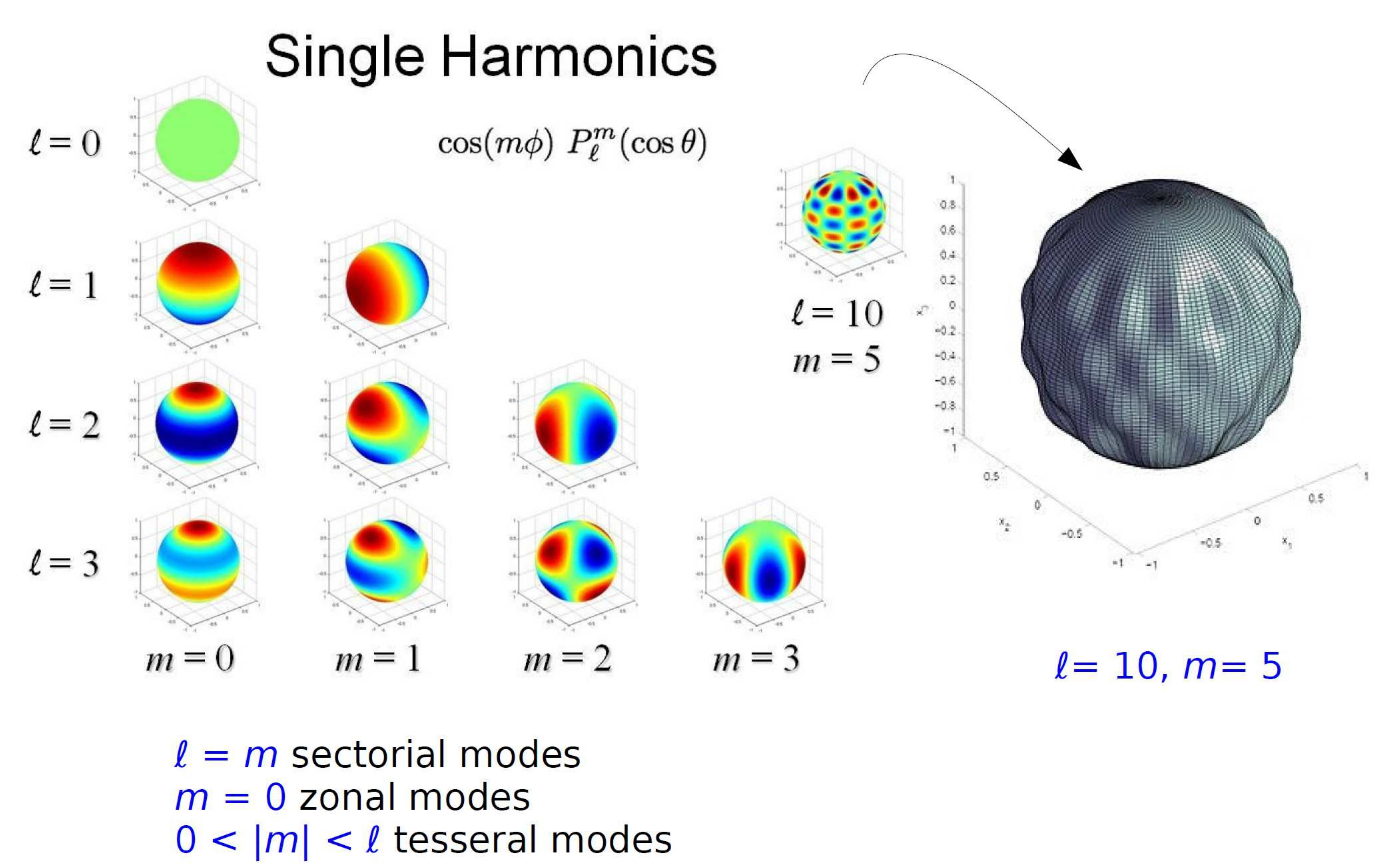}
  \caption{{\it Left:} some examples of single spherical harmonics. 
{\it Right:} The specific case of $\ell= 10, m= 5$ and its 
three-dimensional representation. Figures adapted from
{\tt www.atmos.albany.edu}.}
  \label{fig:2}
\end{figure}

Within the framework of the linear theory of stellar pulsations,
in which the perturbations of the stellar fluid are assumed
to be small, spheroidal nonradial stellar pulsations are described by the
Lagrangian displacement vector (Unno et al. 1989):

\begin{equation}
  \vec{\xi}_{k \ell m}= \left[
                      \xi_r^{k \ell m}(r),
                      \xi_h^{k \ell m}(r) \frac{\partial }{\partial \theta},
                      \xi_h^{k \ell m}(r) \frac{1}{\sin \theta} \frac{\partial }
                         {\partial \phi}\right] Y^m_{\ell} e^{i \sigma_{k \ell m} t}, 
\end{equation}  

\noindent where $Y^m_{\ell}(\theta, \phi)$ are the spherical harmonic
functions, $\sigma_{k \ell m}$ is the pulsation eigenfrequency, and $\xi_r^{k \ell m}(r)$ y
$\xi_h^{k \ell m}(r)$ are the radial and horizontal eigenfunctions, respectively,
of the eigenmode characterized by the set of quantum numbers ${k,
  \ell, m}$. Therefore, in the linear theory of pulsations, the
eigenmodes are described by a sinusoidal temporal dependence given by
the factor $e^{i \sigma_{k \ell m} t}$, an angular dependence defined by
the spherical harmonics $Y^m_{\ell}(\theta, \phi)$, and a radial
dependence through the eigenfunctions $\xi_r^{k \ell m}(r)$ y  $\xi_h^{k \ell m}(r)$, which
have to be determined by numerically solving the set of  differential
equations\footnote{An exception is the case of a self-gravitating
homogeneous ($\rho$= constant) sphere, that admits an analytic
solution (Pekeris 1938), although such a configuration
is not of much astronomical relevance.} that describe the nonradial
pulsations (Unno et al. 1989). This set is a fourth-order system
of equations in real variables in the
adiabatic approximation, and a sixth-order system of equations in
complex variables for the full problem of nonadiabatic pulsations. 
The quantum numbers are defined as: (1) harmonic degree, $\ell = 0, 1, 2, 3,
\cdots, \infty$, that represents $(\ell-m)$ nodal lines (parallel
circles) on the stellar surface; (2) azimuthal order, $m= -\ell,
\cdots, -2, -1, 0 , +1, +2, \cdots, +\ell$, that represents  nodal
lines (meridian circles) on the stellar surface, and (3) radial order,
$k= 0, 1, 2, 3, \cdots, \infty$, that represent nodal concentric
spherical surfaces on which the displacement is null. We note that in
the absence of any physical agent able to
remove spherical symmetry, such as magnetic fields or rotation,
the eigenfrequencies are $(2\ell+1)$-fold degenerate
in $m$. In Fig. \ref{fig:2} we show an illustration of single
spherical harmonics.

There exist three families of spheroidal pulsation modes: (1) $p$ modes:
characterized by large pressure variations and displacements mostly in the radial direction,
the dominant restoring force being the compressibility. $p$ modes have high frequencies
(short periods); (2) $g$ modes: characterized by small pressure variations and
displacements almost tangential, being the dominant restoring force is
the buoyancy. $g$ modes have low frequency (long periods); (3) $f$ modes:
have intermediate characteristics between $p$ and $g$ modes,
and exist only for $\ell > 1$. Generally,
these modes do not have nodes in
the radial direction ($k= 0$), except for stellar models with a high
central density.

The first pulsating WD star, HL Tau 76, was discovered by Landolt (1968).
Pulsating WDs exhibit $g$-mode pulsations with $\ell= 1$ and
$2$\footnote{Modes with higher values of $\ell$ could be excited in WDs,
but their detection should be hampered by geometric
effects of cancellation (Dziembowski 1977).}.
There is only one reported case of a pulsating WD (specifically,
a low-mass WD) that could be pulsating with periods at $\sim 100$ s
associated to $p$ modes or even radial ($\ell= 0$) modes and low
radial order $k$ (Hermes et al. 2013a). The amplitude of the
variations of WD pulsations are typically between 0.005 y 0.4 mag. A plethora of light
curves, some sinusoidal and with small amplitudes, other nonlinear and
with large amplitudes, are observed. Generally, pulsating WDs are multimode
pulsators, that is, they exhibit more than just a single period. In
some cases, the pulsation spectrum contains linear frequency
combinations of genuine pulsation eigenmodes, likely produced by the
outer convection zone of the star.

\begin{figure}[!ht]
  \centering
  \includegraphics[width=0.80\textwidth]{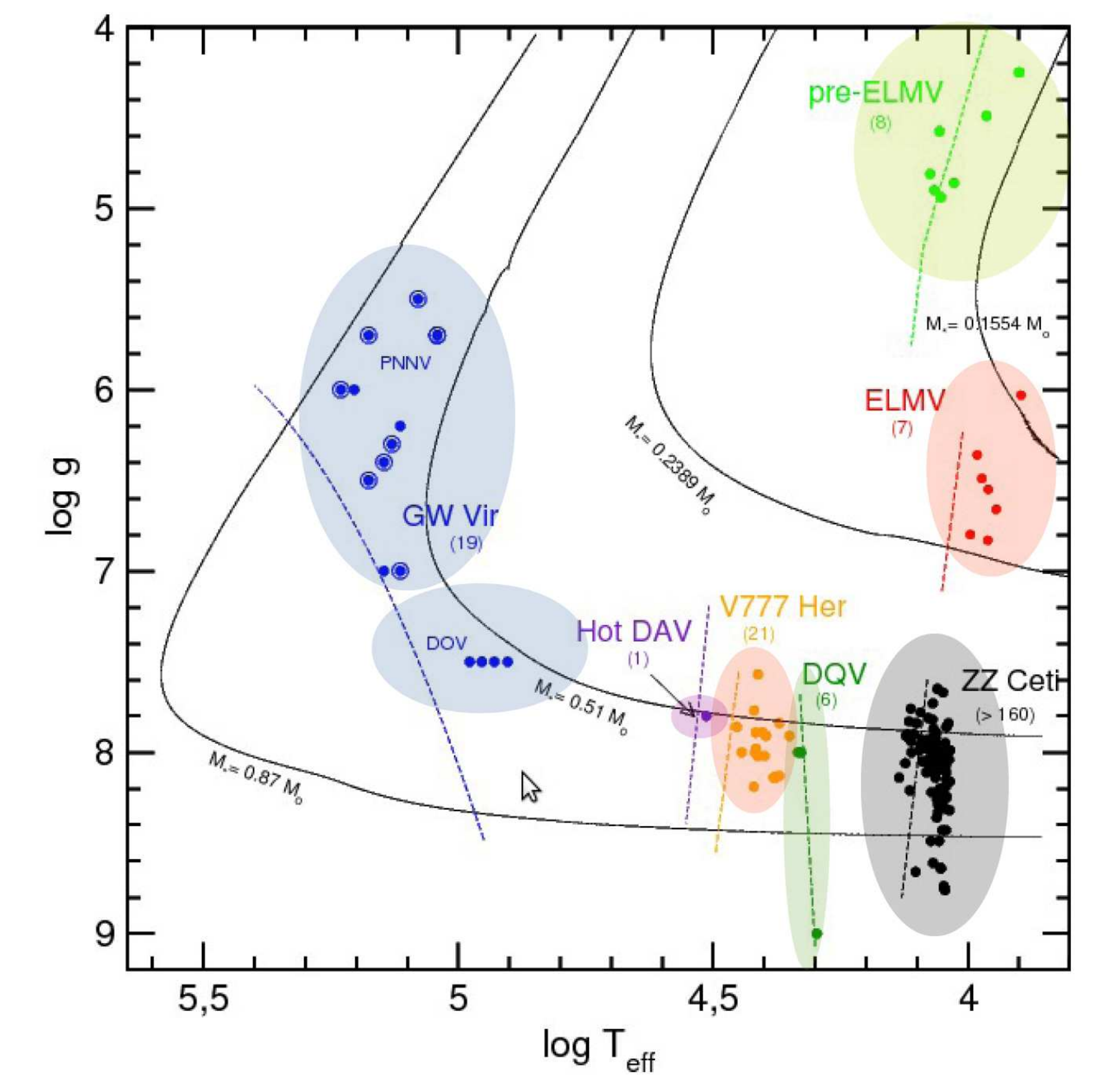}
  \caption{The location of the known classes of pulsating
    WD and pre-WD stars (dots of different colors) in
    the  $\log T_{\rm eff} - \log g$
    plane. We emphasize  in particular the ELMV and pre-ELMV stars
    with big red and light green circles,
    respectively. In parenthesis we include the number of known members of
  each class.  Two post-VLTP (Very Late Thermal Pulse) 
  evolutionary  tracks  for H-deficient WDs and two 
  evolutionary tracks for low-mass He-core WDs  are  plotted   for reference.  
  Also   shown  with dashed lines are  the  theoretical
  blue   edges  of  the different instability domains.} 
  \label{fig:3}
\end{figure}

\section{The zoo of pulsating WDs}

An increasing number of kinds of WD pulsators has been discovered in
the last years. At present, there are seven classes of pulsating WDs.
To begin with, the variables ZZ Ceti or DAVs
(pulsating WDs with almost pure H atmospheres) are the most numerous
ones. The other classes comprise the DQVs (atmospheres rich in He
and C), the variables V777 Her or DBVs (atmospheres almost
pure in He), the Hot DAVs (H-rich atmospheres), and the
variables GW Vir or pulsating PG1159 pre-WD stars (atmospheres dominated
by C, O, and He) that include the DOVs and PNNVs objects. To these
families of pulsating WD stars, we have to add the ELMVs
(extremely low-mass WDs variable) and the pre-ELMVs (the probable
precursors of ELMVs). In Fig. \ref{fig:3} we depict
the location of the several families of pulsating WDs known hitherto in
the $\log T_{\rm eff}- \log g$ plane.

Regarding the driving mechanisms involved in the excitation
of the pulsations in WDs, there is a strong consensus that
they correspond to thermal processes that give place to
self-excited pulsations. The more relevant mechanism is the
$\kappa-\gamma$ mechanism acting in partial ionization regions of
the dominant chemical element: H$_{\rm I-II}$ (DAVs, ELMVs), He$_{\rm I-II}$
(DBVs), He$_{\rm II-III}$ (pre-ELMVs), C$_{\rm V-VI}$ and 
and O$_{\rm VII-VIII}$ (pulsating PG119 pre-WD stars). When
the outer convection zone deepens,
WD pulsations are excited by the ``convective driving''
mechanism (DAVs, DBVs, ELMVs). Finally, the $\varepsilon$  mechanism
due to stable nuclear
burning could be responsible for the excitation of short-period
$g$ modes in GW Vir stars (He burning; C\'orsico et al. 2009),
ELMVs (H burning; C\'orsico et al. 2014a), and in
average-mass DAV WDs evolved from low-metallicity progenitors
(H burning; Camisassa et al. 2016).

In the next section, we concentrate in the last theoretical
results about pre-ELMV and ELMV variable stars. A thorough
description of the pulsation properties of these pulsating WD stars
can be found in the papers by Steinfadt et al.  (2010),
C\'orsico et al.  (2012), Jeffery \& Saio (2013),
Van Grootel  et  al.   (2013),  C\'orsico  \&  Althaus  et  al.
(2014ab, 2016),  C\'orsico  et  al.   (2016),
and Gianninas et al.  (2016). 

\section{Pulsating low-mass He-core WDs}

Low-mass  WDs ($M_{\star} \lesssim 0.45 M_{\odot}$), including ELM WDs
($M_{\star} \lesssim 0.18-0.20 M_{\odot}$, H-rich atmospheres),   which
are currently being detected through the  ELM survey (see Brown et al.
2016 and references therein), likely harbor
cores made of He. They are thought to be the outcome of  strong
mass-loss episodes at the red giant branch stage of low-mass stars in binary
systems  before the He flash onset that, in this way, is avoided
(Althaus et al. 2013; Istrate et al. 2016). Some ELM WDs exhibit
long-period  ($\Pi \sim 1000-6300$ s)  $g$-mode pulsations
(Hermes et al. 2012, 2013ab; Kilic et al. 2015; Bell et al. 2015);
they are called
ELMV pulsating stars. The asteroseismological study of ELMVs
($7000  \lesssim T_{\rm eff} \lesssim 10\,000$ K and $6 \lesssim \log g
\lesssim 7$; red circles in Fig. \ref{fig:3})
can helps to sound their interiors and ultimately to
yield valuable clues about their formation scenarios.
On the other hand, short-period ($\Pi \sim 300-800$ s)
$p$- or even radial-mode ($\ell= 0$)
pulsations in five  objects that are probably precursors of ELM WDs
have been recently discovered
(Maxted et al. 2013, 2014; Gianninas et al. 2016). These stars have
$8000 \lesssim T_{\rm eff} \lesssim 13\,000$ K and $4
\lesssim \log g \lesssim 5.5$ (green circles in Fig. \ref{fig:3}) and
show a mixture of H and He on the surface. They are known as pre-ELMV stars
and constitute the newest class of pulsating WD stars \footnote{There are other
  three stars (Corti et al. 2016, Zhang et al. 2016) that exhibit
  long-period  ($\Pi
\sim 1600-4700$ s) pulsations, located at nearly
the same region of the HR diagram, but their nature as low-mass proto-WDs
cannot be confirmed at the moment; i.e., they could be, alternatively,
$\delta$ Scuti- and/or SX Phe-like pulsators.}.

Below, we describe the predictions of current stability analysis
on pre-ELMVs and ELMVs and how these compare  with the observations. 

\begin{figure}[!ht]
  \centering
  \includegraphics[width=.50\textwidth]{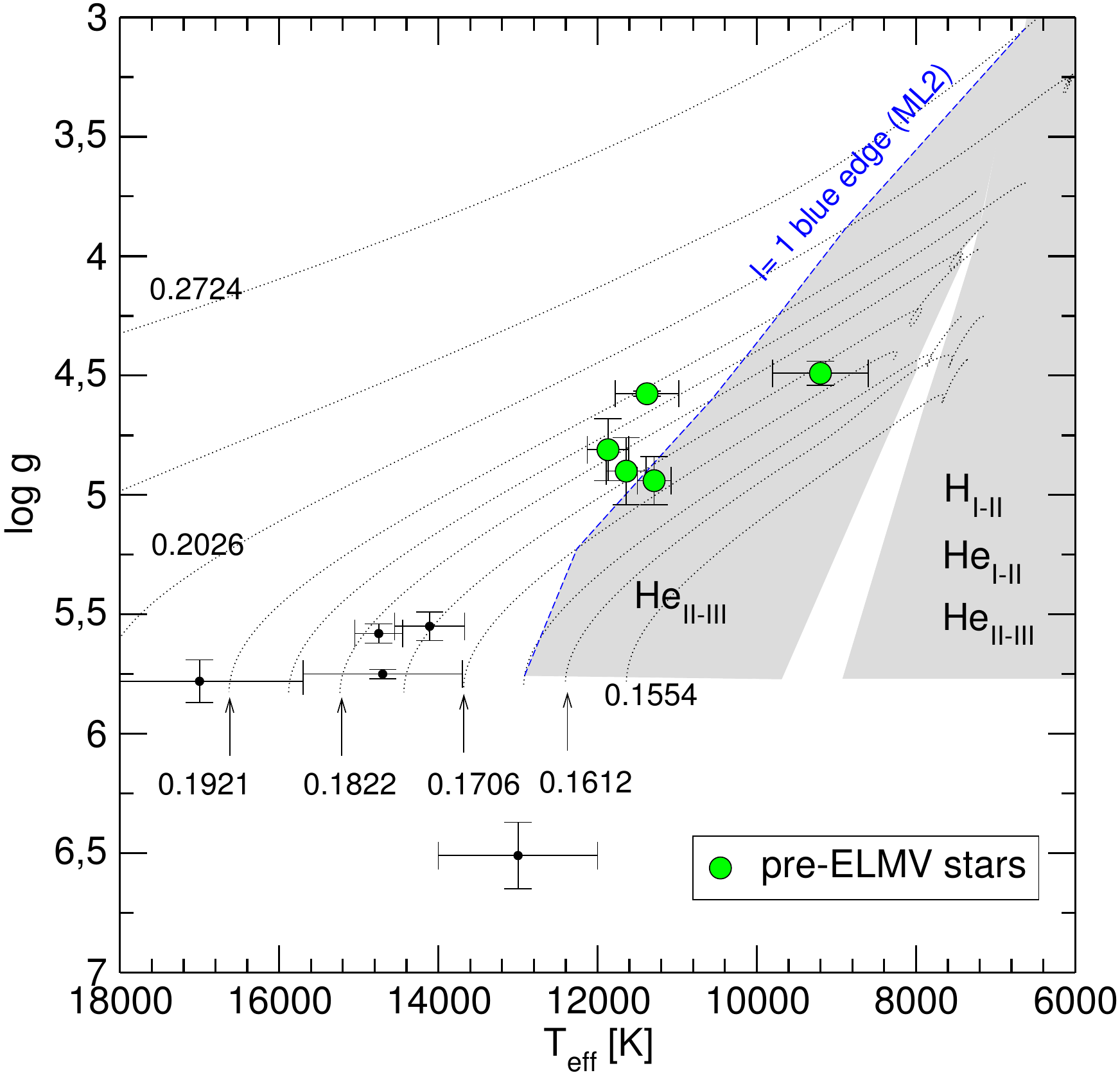}~\hfill
  \includegraphics[width=.50\textwidth]{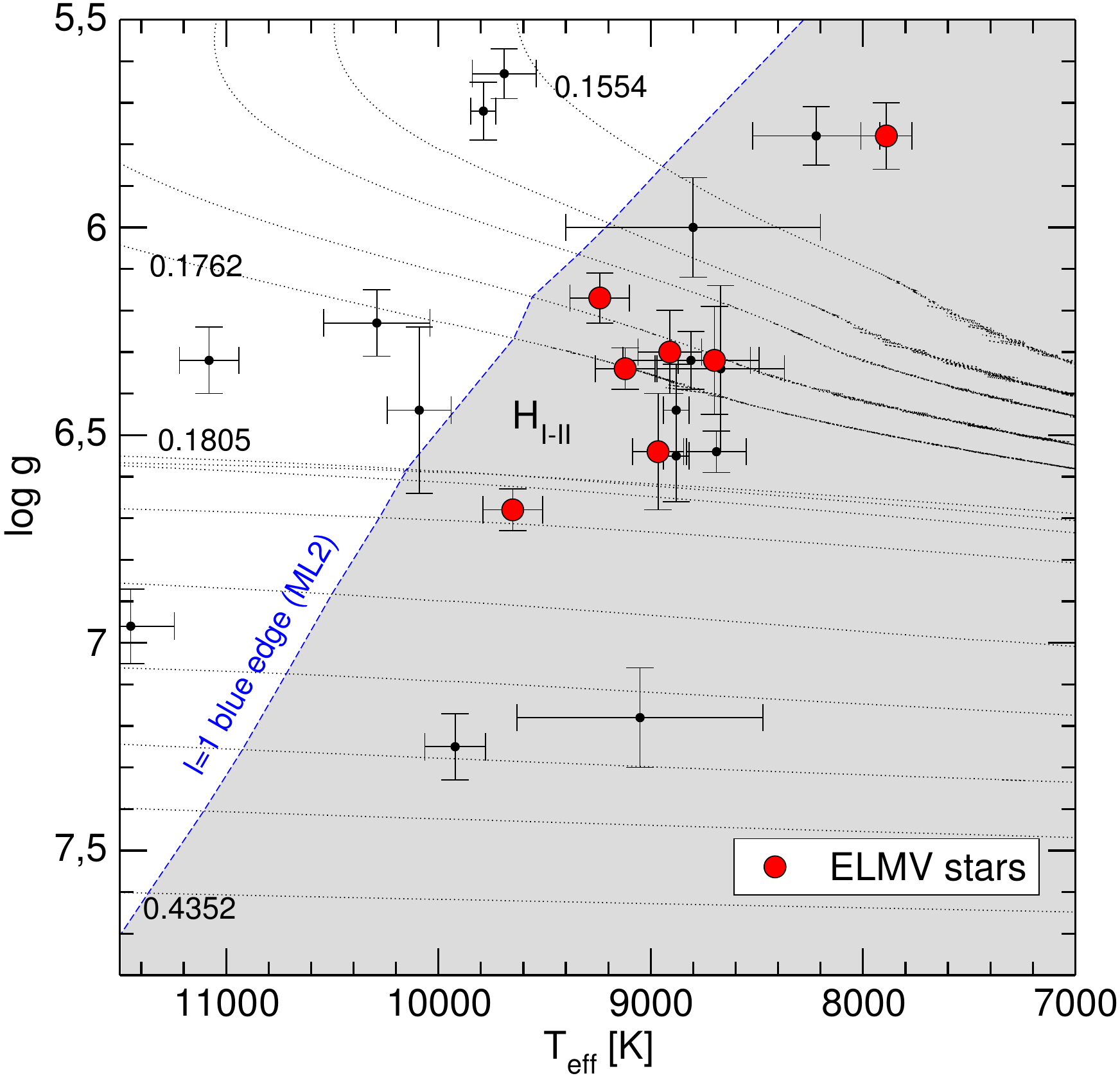}
  \caption{{\it Left:} The
  $T_{\rm eff} - \log g$ diagram showing low-mass He-core  pre-WD
  evolutionary tracks (dotted curves) computed neglecting element
  diffusion. Numbers correspond to the stellar mass of some sequences.
  Green circles with error bars correspond to the confirmed pre-ELMV stars, and
  black dots depict  the location of stars not observed to vary.
  The dashed blue line indicates the nonradial dipole ($\ell= 1$)  blue edge of
  the pre-ELMV instability domain  (emphasized as a gray area)
  of $p$ and radial modes  due to the $\kappa-\gamma$ mechanism
  acting at the He$_{\rm II-III}$   (at high $T_{\rm eff}$) and
  He$_{\rm I-II}$/H$_{\rm I-II}$ (at low $T_{\rm eff}$) 
  partial ionization   regions,  as obtained by C\'orsico et
  al. (2016). \protect {\it Right:} The same diagram but for low-mass
  He-core WD evolutionary tracks (final cooling branches).  The
  locations of the seven known ELMVs are marked with red circles
  ($T_{\rm eff}$ and $\log g$ computed with 1D model atmospheres after
  3D corrections).  The gray region bounded by the dashed blue line
  corresponds to the instability  domain of $\ell= 1$ $g$ modes due to
  the $\kappa-\gamma$  mechanism acting at the H$_{\rm I-II}$ partial
  ionization region,  according to C\'orsico \& Althaus (2016).}
  \label{fig:4}
\end{figure}

\subsection{Pre-ELMVs}

Here, we describe the main nonadiabatic pulsation
results of the study by C\'orsico et al.  (2016). The stability properties
of He-core, low-mass pre-WD models extracted from the computations
of Althaus et al. (2013),  calculated assuming the ML2
prescription for the MLT theory of convection (Tassoul et al. 1990), 
covering a range of
effective temperatures of $25\,000\ {\rm K} \gtrsim T_{\rm eff}
\gtrsim 6000$ K and a range of stellar masses of $0.1554 \lesssim
M_{\star}/M_{\odot} \lesssim 0.2724$, were analyzed. Gravitational settling
was neglected. For each model, the pulsational stability of
radial $(\ell= 0)$, and nonradial $(\ell= 1, 2)$ $p$ and $g$ modes
with periods in the range $10\ {\rm s} \lesssim \Pi \lesssim 20\,000$
s was assessed. The results are shown in the left panel of Fig.
\ref{fig:4}, in which we depict a spectroscopic HR
diagram (the $T_{\rm eff} - \log g$ plane) that includes
low-mass He-core pre-WD evolutionary tracks (dotted curves)
along with the location of the
known pre-ELMV stars (green circles) and
stars not observed to vary (black dots).  
The theoretical blue edge of the dipole ($\ell= 1$)
pre-ELMV instability domain (gray area) due to the $\kappa-\gamma$
mechanism acting at the He$_{\rm II-III}$
($\log T \sim 4.7$) and He$_{\rm I-II}$/H$_{\rm I-II}$ ($\log T \sim
4.42$/$\log T \sim 4.15$) 
partial ionization   regions is displayed with a 
dashed blue line. These results are in excellent  
agreement with the predictions of the stability analysis carried 
out by Jeffery \& Saio (2013). The location of the
theoretical blue edge is almost insensitive to the value of $\ell$ and
the prescription for the MLT theory of convection
(that is, the efficiency of convection) adopted in the construction of
the stellar models. The blue edge of radial modes is
$\sim 1000$ K cooler than for nonradial modes. 
Our computations roughly account for the location
of the known pre-ELMVs, although the theoretical blue edge
should be $\sim 900$ K hotter in order to achieve a better agreement. 
This can be accomplished with higher He abundances at the
driving region. This issue should be investigated in more detail.

A crucial aspect of these computations is that they neglect the effect
of gravitational settling. When we take into account gravitational
settling, the driving regions quickly becomes depleted in He, and the
instability region in the $T_{\rm eff}-\log g$ diagram shrinks,
leading us to the inability to explain the existence of \emph{any} of these
pulsating stars.  This indicates that \emph{gravitational settling
  could not be operative, or could be slowed, in the pre-WD
  stage}. 

\subsection{ELMVs}

For the study of ELMVs, stellar models of He-core,
low-mass WDs extracted from the computations of Althaus et al. (2013),
having H-pure atmospheres and taking into account  three different prescriptions for
the  MLT theory of convection
(ML1, ML2, ML3; see Tassoul et al. 1990 for their
definitions), covering a range
of effective  temperatures of   $13\,000\ {\rm K} \lesssim  T_{\rm
  eff} \lesssim 6\,000$ K and a range of stellar masses of $0.1554
\lesssim M_{\star}/M_{\odot} \lesssim 0.4352$, were considered.
For each model, the pulsation stability of radial ($\ell= 0$) and nonradial
($\ell= 1, 2$) $p$ and $g$ modes with periods  from a range $10\ {\rm
  s} \lesssim \Pi \lesssim 18\,000$ s for the sequence with $M_{\star}= 0.1554
M_{\odot}$, up to a range of periods of $0.3\ {\rm s} \lesssim \Pi \lesssim
5\,000$ s  for the sequence of with $M_{\star}= 0.4352 M_{\odot}$ was assessed.
Complete details of these calculations are given in C\'orsico \& Althaus (2016).
The results are shown in the right  panel of Fig. \ref{fig:4},
in which we depict the spectroscopic HR diagram including the low-mass
He-core WD evolutionary tracks (final cooling
branches),  along with the location of the seven known ELMVs (red
circles), where  $T_{\rm eff}$ and $\log g$ have been computed with 1D
model atmospheres  after 3D corrections.  The instability domain of
$\ell= 1$ $g$ modes due to the $\kappa-\gamma$  mechanism acting at
the H$_{\rm I-II}$ ($\log T \sim 4.15$)  partial ionization region
is emphasized  with a gray region bounded by a dashed
blue line corresponding  to the blue edge of the  instability
domain. Our results are in good agreement with those of  C\'orsico et
al. (2012) and  Van Grootel et al. (2013). Some short-period $g$ modes
are destabilized mainly  by the $\varepsilon$ mechanism due to stable
nuclear burning at the basis of the  H envelope, particularly for
model  sequences with $M_{\star} \lesssim 0.18 M_{\odot}$ (see
C\'orsico \& Althaus 2014a for details).  The blue edge of the
instability domain is hotter for
higher stellar mass and larger convective efficiency.  The ML2 and ML3
versions of the MLT theory of convection are the only ones that
correctly account for the location of the seven known ELMV
stars. We found a weak sensitivity of the blue edge of $g$ modes
with $\ell$, and the blue edges corresponding
to radial and nonradial $p$ modes are  somewhat ($\sim 200$ K)  hotter
than the blue edges of $g$ modes. 

\section{Conclusions}

WD asteroseismology is consolidating as one of the most attractive
avenues to explore the properties of WDs. It can provide
crucial information about the internal structure of these stars,
which leads to a detailed knowledge of evolutionary processes
experienced by their parent stars. It is expected that in the coming
years, unparalleled progress in the asteroseismic study
of the origin and evolution of WDs
will be reached with the help of space missions such as TESS
(Transiting Exoplanet Survey Satellite, {\tt https://tess.gsfc.nasa.gov/}).

\agradecimientos

I want to warmly thank the members of the LOC and
SOC of the ``Terceras Jornadas de Astrof\'isica Estelar'' for the
valuable work that led to the realization of this excellent scientific
meeting.

\begin{referencias}
                                                                                
\reference Althaus, L. G., Miller Bertolami, M. M., \& C\'orsico, A. H. 2013, \aap, 557, A19
\reference Althaus, L. G., C\'orsico, A. H., Isern, J., \& Garc\'ia-Berro, E. 2010, \aapr, 18, 471
\reference Bell, K. J., Kepler, S. O., Montgomery, M. H., et al. 2015, 19th European Workshop on
White Dwarfs, 493, 217
\reference Brown, W. R., Gianninas, A., Kilic, M., Kenyon, S. J., \& Allende Prieto, C. 2016, \apj,
818, 155
\reference Camisassa, M.~E., C{\'o}rsico, A.~H., Althaus, L.~G., \& Shibahashi, H.\
2016, arXiv:1606.04367 
\reference Catal{\'a}n, S., Isern, J., Garc{\'{\i}}a-Berro, E., \& Ribas, I.\ 2008, \mnras, 387, 1693
\reference Catelan, M., \& Smith, H.~A.\ 2015, Pulsating Stars (Wiley-VCH), 2015,  
\reference Chandrasekhar, S. 1939, Chicago, Ill., The University of Chicago press
\reference C\'orsico, A. H., Althaus, L. G., Serenelli, A. M., et al. 2016, \aap, 588, A74
\reference C\'orsico, A. H., \& Althaus, L. G. 2016, \aap, 585, A1
\reference C\'orsico, A. H., \& Althaus, L. G. 2014b, \aap, 569, A106
\reference C\'orsico, A. H., \& Althaus, L. G. 2014a, \apj, 793, L17
\reference C\'orsico, A. H., Romero, A. D., Althaus, L. G., \& Hermes, J. J. 2012, \aap, 547, A96
\reference C\'orsico, A. H., Althaus, L.~G., Miller Bertolami, M.~M., Gonz{\'a}lez P{\'e}rez, J.~M., \& Kepler, S.~O.\ 2009, \apj, 701, 1008 
\reference Corti, M. A., Kanaan, A., C\'orsico, A. H., et al. 2016, \aap, 587, L5
\reference Dziembowski, W.\ 1977, Acta Astron., 27, 203 
\reference Fontaine, G., \& Brassard, P. 2008, \pasp, 120, 1043
\reference Garc{\'{\i}}a-Berro, E., Torres, S., Althaus, L.~G., et al.\ 2010, \nat, 465, 194 
\reference Gianninas, A., Curd, B., Fontaine, G., Brown, W. R., \& Kilic, M. 2016, \apj, 822, L27
\reference Hansen, B.~M.~S., Kalirai, J.~S., Anderson, J., et al.\ 2013, \nat, 500, 51 
\reference Harris, H.~C., Munn, J.~A., Kilic, M., et al.\ 2006, \aj, 131, 571 
\reference Hermes, J. J., Montgomery, M. H., Gianninas, A., et al. 2013a, \mnras, 436, 3573
\reference Hermes, J. J., Montgomery, M. H., Winget, D. E., et al. 2013b, \apj, 765, 102
\reference Hermes, J. J., Montgomery, M. H., Winget, D. E., et al. 2012, \apj, 750, L28
\reference Isern, J., Garc{\'{\i}}a-Berro, E., Hernanz, M., Mochkovitch, R., \& Torres, S.\ 1998, \apj, 503, 239 
\reference Istrate, A., Marchant, P., Tauris, T. M., et al. 2016, arXiv:1606.04947
\reference Jeffery, C. S., \& Saio, H. 2013, \mnras, 435, 885
\reference Kepler, S.~O., Koester, D., \& Ourique, G.  2016a, Science, 352, 67 
\reference Kepler, S.~O., Pelisoli, I., Koester, D., et al.\ 2016, \mnras, 455, 3413 
\reference Kepler, S.~O., Pelisoli, I., Koester, D., et al.\ 2015, \mnras, 446, 4078 
\reference Kilic, M., Hermes, J. J., Gianninas, A., \& Brown, W. R. 2015, \mnras, 446, L26
\reference Kleinman, S.~J., Kepler, S.~O., Koester, D., et al.\ 2013, \apjs, 204, 5 
\reference Landolt, A.~U.\ 1968, \apj, 153, 151 
\reference Maxted, P. F. L., Serenelli, A. M., Marsh, T. R., et al. 2014, \mnras, 444, 208
\reference Maxted, P. F. L., Serenelli, A. M., Miglio, A., et al. 2013, \nat, 498, 463
\reference McCook, G.~P., \& Sion, E.~M.\ 1999, \apjs, 121, 1 
\reference Mestel, L.  1952, \mnras, 112, 583 
\reference Pekeris, C.~L.\ 1938, \apj, 88, 189 
\reference Steinfadt, J. D. R., Bildsten, L., \& Arras, P. 2010, \apj, 718, 441
\reference Tassoul, M., Fontaine, G., \& Winget, D. E. 1990, \apjs, 72, 335
\reference Unno, W., Osaki, Y., Ando, H., Saio, H., \& Shibahashi, H.\ 1989, Nonradial oscillations of stars, Tokyo: University of Tokyo Press, 1989, 2nd ed.,  
\reference Van Grootel, V., Fontaine, G., Brassard, P., \& Dupret, M.-A. 2013, \apj, 762, 57
\reference Winget, D. E., \& Kepler, S. O. 2008, \araa, 46, 157
\reference Woosley, S. E., \& Heger, A. 2015, \apj, 810, 34
\reference York, D.~G., Adelman, J., Anderson, J.~E., Jr., et al.\ 2000, \aj, 120, 1579 
\reference Zhang, X. B., Fu, J. N., Li, Y., Ren, A. B., \& Luo, C. Q. 2016, \apj, 821, L32
\end{referencias}

\end{document}